\documentclass{article}
\usepackage[english]{babel}
\usepackage[T2A]{fontenc}
\usepackage[utf8]{inputenc}
\usepackage{mathtext}
\usepackage{graphicx}
\usepackage{amsfonts}
\usepackage{amssymb}
\usepackage{amsmath}
\usepackage{float}
\usepackage{indentfirst}
\usepackage[a4paper,margin=2.5cm]{geometry}
\begin{document}
\title{Dark states of atomic ensembles}

\author{Yuri I. Ozhigov\thanks{Lomonosov Moscow State University, Faculty of Computational Mathematics and Cybernetics, Moscow, Russia}\hspace{0.15cm}, Nikita A. Skovoroda\thanks{The same place}
}

\maketitle

\begin{abstract}
Ensemble of identical two level atoms in dark state neither adsorbs nor emits photons due to destructive interference. It can be used for the source of energy for nano-devices. In Tavis-Cummings  cavity the change of light-atom coupling cannot destroy the dark state. We propose the method of dark states preparation based on photon pumping and drain from the optical cavity with two atoms, one from which has Stark or Zeeman splitting of energy levels. This splitting is removed before photon drain from the cavity. The dark state yield after one such cycle has the order of energy level splitting. We show the scheme of the experiment, in which such cycles repeat until the dark state is produced with high probability and the results of computer simulation. 

\vspace{0.2cm}

\end{abstract}


\section{Introduction and background}

The optical cavity - a remarkable device that uses the interaction of light and matter, as they are able to concentrate single photons around the compact atomic ensembles. They allow to study such important for applications phenomena as conductivity of atomic excitations in the presence of noise (\cite{HP}), photon echo and quantum photonic memory (\cite{Moi}), quantum computations on nonlinear optical processes (\cite{A}), etc.

An important type of quantum states of field and matter that can be obtained and studied in optical cavities - the states of atomic ensemble that can not absorb or emit light (see, for example, \cite{FB}); such states can be used as a broad class of quantum switches in microelectronic devices and as the source of energy for nano-devices. The example is the dipole blockade of Rydberg atoms, which mechanism is based on the change of atomic frequency (see, for example, \cite{BAC}). Dark states of multilevel ensembles arise due to jump on the energy level, transitions from which are forbidden. Such states were treated in Zeeman-degenerate atomic ensembles (\cite{BB}), dark state polaritons - form-stable coupled excitations of light and
matter were considered in \cite{FL}.  

We study dark states of ensembles of $n$ two level atoms in optical cavity in Jaynes-Tavis-Cummings or Dick model(see \cite{JC}, \cite{FB}). We call a state dark if it can neigther absorb, nor emit the photon.  Such states must be among eigen states of JTC Hamiltonian
\begin{equation}
H_{JTC}=h\omega_ca^+a+\sum\limits_{i=1}^nh\omega_i\sigma_i^+\sigma_i+
\sum\limits_{i=1}^nhg_i(\sigma_i^++\sigma_i)(a^++a),
\label{JTC}
\end{equation}
where atomic frequencies $\omega_i$ may be different, as well as field-atom interaction constants $g_i$ so that all detunings are relatively small: $d_i=\omega_c-\omega_i,\ |d_i|\ll\omega_c$. For the relatively small interaction: $g_i/\omega_c\ll 1$ we can use rotating wave approximation (RWA) that is to omit not conserving energy summands $a^+\sigma_j^+,\ a\sigma_j$, which in the interaction picture represent rapidly oscillating summands. Interaction summand thus acquires the form $h\sum\limits_{n=1}^ng_i(a^+\sigma_j+a\sigma^+_j)$ where we can assume that $g_j$ are real (to make them real we have to change the phase of basic states). In what follows we use RWA approximation without special mentioning, though our conclusions does not substantially depend on the precision of RWA. 

We denote by $|0\rangle$ and $|1\rangle$ the ground and excited state of each atom. Let us divide the set of atoms to $n/2$ pairs ($n$ even) and use lower index $j$ to indicate the number of the pair, $j=1,2,\ldots,n/2$. There exists the series of dark states of the form 
\begin{equation}
\frac{1}{2^{n/4}}\bigotimes\limits_{j=1}^{n/2}(|01\rangle_j-|10\rangle_j).
\end{equation}
In this state atoms of each pair behave like Buridan donkey: any attempt to emit a photon from one atom is cancelled by the anty-symmetric attempt from the other. The same is right for the absorption of photons. Hence, atoms in dark pair interact neither with other atoms, nor with photons. 

Can we get a dark state as a result of the unitary evolution with adiabatically changing Hamiltonian, which includes only the interaction between light and matter? Physically allowable time reversal in this evolution would lead to the disintegration of the dark state that is impossible. Thus, the dark state preparation requires  non-adiabatic transition. That transition can be a) fast mechanical movement of atoms (for example, as a result of nuclear transformations), which leads to a sharp change in the Hamiltonian, or b) a rapid change in the Hamiltonian parameters due to a sharp turn on or off the electric or magnetic field.
We will show the reality of the second way, which will be used by Zeeman-Stark (ZS) effect in which the constant external electric or magnetic field shifts the energy levels of atoms.

\section{Dark states stability}

Hamiltonian (\ref{JTC}) concerns the case where atom-light interaction is the same for the different atoms. But this is not always the case. Since the strength of the photon electric field inside the cavity has a sinusoidal shape $E(x)=E_0 sin(\omega_c x/c)$, where the length of the cavity $L=\pi c/\omega_c$ is the half of photon wave length, $x$ the cordinate of the atom along the axis of the cavity, $E_0=\sqrt{h\omega_c/2\varepsilon_0V}$, $V$ is the volume of cavity, $\varepsilon_0$ is electric constant, atom-light interaction $d_a\cdot E(x)$ ($d_a$ - matrix element of the dipole momentum of the atomic transition $|0\rangle\rightarrow |1\rangle$) depends on the position of the atom inside the cavity. 

Use lower index $p$ for photon, $a$ for atoms. We denote by upper tilde the division of any Hamiltonian to Plank constant: $\tilde H=H/h$. The restriction of $\tilde H_{JTC}$ to the subspace, spanned by states $|0\rangle_p|10\rangle_a,\ |0\rangle_p|01\rangle_a, \ |1\rangle_p|00\rangle_a$, has the form
\begin{equation}
H_1(\omega_a^1, \omega_a^2, g_1,g_2)=\left(
\begin{array}{llll}
&\omega_a^1&0&g_1\\
&0&\omega_a^2&g_2\\
&g_1&g_2&\omega_c
\end{array}
\right)
\label{smallHam}
\end{equation}
where $\omega_a^i,\ g_i,\ i=1,2$ denote the frequency of $i$-th atom and the intensity of its interaction with the field. We can make frequencies different for identical atoms by electric voltage resulted from ZS effect. 
Without ZS effect $\omega_a^1=\omega_a^2$ and dark states has the form 
\begin{equation}
|d.s.(g_1,g_2)\rangle=-g_2|10\rangle_a+g_1|01\rangle_a.
\label{noStark}
\end{equation}
Given the dark state of the form (\ref{noStark}) we could easily obtain the state of the form $\frac{1}{\sqrt{2}}(|01\rangle-|10\rangle )$ by the adiabatic change of the Hamiltonian that is possible to do by slow moving of atoms one to the other by optical tweezers. Such change of the atomic state cannot create the dark state because manipulation with optical  tweezers cannot give the required abruptness: the change of Hamiltonian must be faster than the period of Rabi oscillations. 

We introduce the notations $S=\sqrt{4g_1^2+4g_2^2+d^2},\ d=\omega_c-\omega_a,$ the eigenvalues of $H_1$ will then be $\omega_c,\ \frac{1}{2}(\omega_c+\omega_a-S),\ \frac{1}{2}(\omega_c+\omega_a+S)$ and the corresponding eigenstates will be
\begin{equation}
\{ -g_2,g_1,0\},\ \{ -\frac{2 g_1}{\omega_a-\omega_c+S},-\frac{2 g_2}{\omega_a-\omega_c+S},1\},\ 
\{ \frac{2 g_1}{\omega_c-\omega_a+S},\frac{2 g_2}{\omega_c-\omega_a+S},1\}
\label{noStarkStates}
\end{equation}

In the case of ZS shift we introduce the notations 
$$
A = -(\omega_c+\omega_a^1+\omega_a^2),\ 
B=\omega_c\omega_a^1+\omega_c\omega_a^2+\omega_a^1\omega_a^2-g_1^2-g_2^2,\
C=g_1^2\omega_a^2+g_2^2\omega_a^1-\omega_c\omega_a^1\omega_a^2
$$
and let $\beta_1,\ \beta_2,\ \beta_3$ be roots of the polynomial $x^3+Ax^2+Bx+C$. Eigenvalues of Hamiltonian $H_1$ of the form  (\ref{smallHam}) thus have the form $\beta_1,\ \beta_2,\ \beta_3$ and the corresponding eigen states will have the form 
\begin{equation}
\{ -\frac{\omega_c-\alpha }{g_1}+\frac{g_2^2}{g_1(\omega_a^2-\alpha)}, \frac{g_2}{\omega_a^2-\alpha},1\}
\label{eigenStates}
\end{equation}
for $\alpha = \beta_1,\beta_2,\beta_3$ correspondingly. The equation (\ref{eigenStates}) is not applicable to the case $\omega_a^1=\omega_a^2$, e.g. without ZS effect because $\omega_a^1,\omega_a^2$ will then be among $\beta_1,\beta_2,\beta_3$ and the denominator in (\ref{eigenStates}) is zero. Hence in the case of $\omega_a^1=\omega_a^2$ we must use the form (\ref{noStarkStates}) for eigen states. 

We see that there is no dark states in the case of nonzero ZS shift. This can be used for dark states fabrication through controlled ZS shift that is possible by switching on and off the voltage for the separate atom, provided the other atom has the different location.

\section{Dark state preparation}

To prepare the dark state we start from the state of the form $|\psi (0)\rangle=|1\rangle_p|00\rangle_a$ with one photon and two atoms in ground states; where the frequency of the first atom is shifted by Stark effect, which we ensure by the proper voltage. To do this we must place atows in the different locations along the central axis of cavity so that their coupling constants with the field will be different. This means that atomic ground states $|0\rangle_a^1$ and $|0\rangle_a^2$ have the different senses: it differ by location, by coefficients $g_1, g_2$, and by frequencies $\omega_a^1, \ \omega_a^2$ of transitions $|0\rangle\leftrightarrow |1\rangle$. 

We call ZS jump the following process of the evolution $|\psi (t)\rangle$. We wait some time $\Delta t$, which is choosen at random, and then abruptly turn off the voltage, which supported ZS shift. Let  frequencies and coupling constant for our atoms without ZS look as $\omega, \omega, g_1,g_2$, and ZS applied to the first atom results in the following values for these parameters: $\omega +ds, \omega, g_1+dg,g_2$, e.g. it slightly changes the parameters of the first atom. We denote $H_1(\omega +ds, \omega, g_1+dg,g_2)$ by $H_{ZS}(ds,dg)$.

 \begin{figure}
\includegraphics[width=0.97\textwidth]{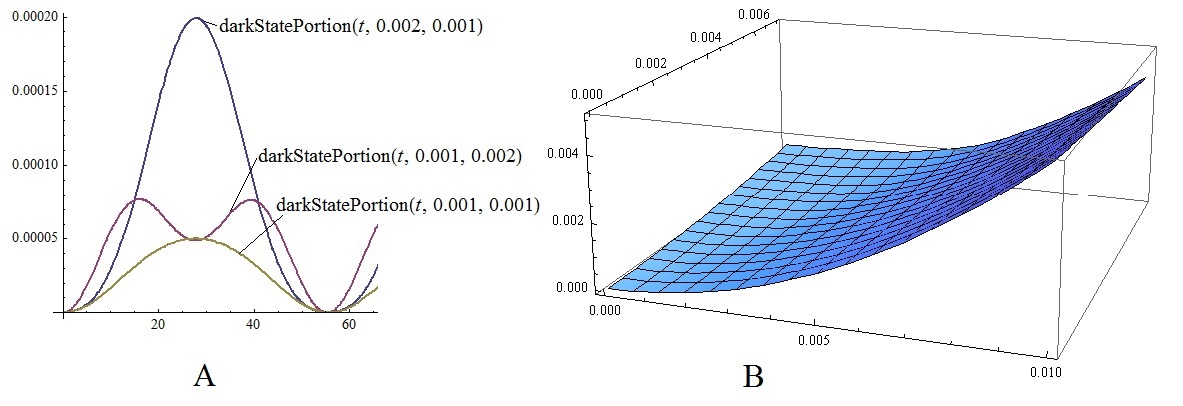}
 \caption{A. The portion of dark states depending on time (numerical simulation); the second and third arguments are dimensionless magnitudes $\tilde ds=ds/\omega_c$ and $\tilde dg=dg/\omega_c$ correspondingly. 
B. Maximal (for all $t$) value of dark state portion as the function of $\tilde ds\in [ 0, 0.01],\  \tilde dg\in [0, 0.007]$. Here the dimensionless time is $t_{physical}/\omega_c$ and we assume that $g_2=g_1/2$.} 
\end{figure}

\begin{figure}
\centering
\includegraphics[width=1.1\textwidth]{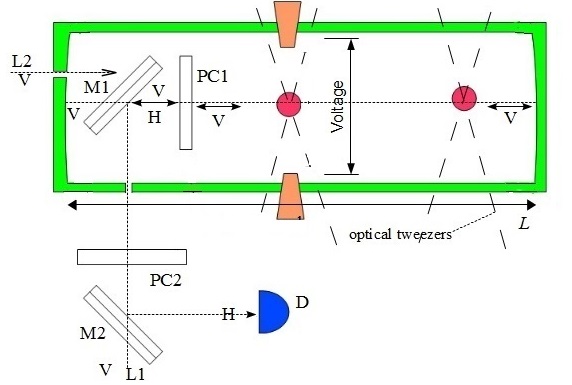}\includegraphics[width=1.1\textwidth]{resonator1.jpg}
 \caption{Preparation of dark singlet in optical cavity. Initially, Pockels cells PC1 and PC2 are both switched on. Photon flies from Laser L1 with vertical pollarization $V$, changes polarization to $H$ after PC2, reflects from the mirror M1 and change its pollarization to $V$ after PC1. Then PC1 must be switched off before photon comes back reflecting from the right wall of cavity. Photon will then locked inside the cavity. After arbitrary time frame PC1 switches on again and photon change the $V$ polarization back to $H$, passes through PC2, which is switched off, reflects from M2 and comes to detector $D$. Alternative way: photon comes from laser L2 with $V$ pollarization and becomes locked in the cavity, then PC1 switches on, and photon comes to detector; PC2 is not needed.}  

\end{figure}

The amplitude of dark state, which appears after ZS jump will be then 
\begin{equation}
\lambda_{ds}=\langle d.s.(g_1,g_2)|\psi (\Delta t)\rangle,\ |\psi(t)\rangle=exp(-\frac{i}{h}H_{ZS}((ds,dg)),
\label{darkAmp}
\end{equation}
and the portion of dark states we yield by ZS jump will be $p_{ds}=|\lambda_{ds}|^2$. Fig.1 shows the behavior of $p_{ds}$. We see that for the reasonable values of $ds, dg$ the portion of dark state is about $10^{-4}$. The valuable ZS shift can be made by the available voltage (see, for example, \cite{Gi}). After switching off voltage in arbitrary moment we swith on Pockels cell PC1 and let the explicit photon fly to the detector. It can be seen from Fig 1, that the probability of finding dark state is small but not vanishing: we must take some medium time instant, that is the dark state portion will be about a half of value from the Fig. 1B. The indicator of obtaining dark state is the absence of photon in detector. The difficulties comes from the errors in laser and detector, and technical limitations of cavities and linear optics elements which serve for the control. For example, the weak point in the first way of state preparation (by laser L1) is the big time of switching on Pockels cell (see the details in \cite{A}). The other drawbacks can come from the photon leakage through the cavity walls. The influence of such factors yet should be estimated. However, such difficulties seem mostly technical. We can move dark pair of atoms in the location where the ZS shift is vanishing by optical tweezers and accumulate this pairs of atoms for the further usage; such a storage will not be corrupted by the process of new dark pair preparation; the new atoms for this must be delivered to the cavity in the ground state.

\section{Conclusion}

We propose the mehod of dark state creation in the optical cavity through non adiabatic evolution with sharply switched off voltage or magnetic field, which causes Stark of Zeeman shift of atomic frequency. The failed  measurement of photon flying on the cavity serves as the verification of dark state presence. The outcome of dark states varies depending on the frequency shift value and can reach about $0.01\%$ with the available shift value. The practical feasibility of our method also depends on the quality of cavity and the speed of Pockels cell switching off that can decrease the outcome rate. Dark state been prepared can be stored outcide of the cavity provided we keep two atoms together by optical tweezers. Dark state can be used as the source of Fock states of photons from example, as one photon source, and also as the source of energy for nano devices.

\end{document}